# Site-specific surface atom valence band structure via X-ray standing wave excited photoemission

Yanna Chen,[1] Leighton O. Jones,[2] Tien-Lin Lee,[3] Anusheela Das,[1] Martín A. Mosquera,[4] Denis T. Keane,[1] George C. Schatz,[2] Michael J. Bedzyk[1,5,*]

[1] Department of Materials Science and Engineering, Northwestern University, Illinois 60208, USA
[2] Department of Chemistry, Northwestern University, Evanston, Illinois 60208, USA
[3] Diamond Light Source, Harwell Science and Innovation Campus, Didcot, OX11 0DE, UK
[4] Department of Chemistry and Biochemistry, Montana State University, Bozeman, Montana 59717, USA
[5] Department of Physics and Astronomy, Northwestern University, Illinois 60208, USA

X-ray standing wave (XSW) excited photoelectron emission was used to measure the site-specific valence band (VB) for ½ monolayer (ML) Pt grown on a SrTiO₃ (001) surface. The XSW induced modulations in the core level (CL) and VB photoemission from the surface and substrate atoms were monitored for three *hkl* substrate Bragg reflections. The XSW CL analysis shows the Pt to have an fcc-like cube-on-cube epitaxy with the substrate. The XSW VB information compares well to a density functional theory calculated projected density of states from the surface and substrate atoms. Overall, this work represents a novel method for determining the contribution to the density of states by valence electrons from specific atomic surface sites.

PACS numbers: 68.49.Uv, 68.47.Jn, 73.20.-r, 79.60.-i

Noble metal/oxide interfaces have important applications in both chemical and physical processes. Dramatic catalytic enhancement is found for low-coverage noble metals supported on oxide surfaces.[1–4] Pt on SrTiO$_3$ is one such case with importance to water-splitting photocatalysis.[1, 3] The Pt/ SrTiO$_3$ interface is also under consideration in thin film electronics for use in resistive random access memory devices.[5] As Pt layers are reduced down to several atomic layers in nanocrystals a larger fraction of the Pt atoms are at the interface which strongly influences catalytic properties.[6] For the Pt/support interface Pt atoms can interact with the TiO$_2$ support surface and withdraw electrons.[7, 8] During a gas catalysis process, the pathway of the catalyzed reaction is governed by the atomic and electronic structures of the catalyst and the interface of catalyst/support.[9] While much research focuses on the Pt catalyst application, there are still open questions over the atomic and electronic properties, especially at the catalyst/support interface.

As a support for Pt, SrTiO$_3$ (STO) has a special morphological preference, since its cubic-P unit cell lattice constant ($a_{STO}$ = 3.905 Å) is a close match to that of the face-centered-cubic (*fcc*) Pt lattice ($a_{Pt}$ = 3.924 Å); and thus beneficial for obtaining an epitaxial interface. In previous work, *fcc* Pt nanocrystals on a STO (001) support were grown by molecular beam epitaxy (MBE) and x-ray standing wave (XSW) excited Pt L$\alpha$ x-ray fluorescence (XRF) was used to find the different interfacial structures dependent on the Pt submonolayer coverage.[10] Using density functional theory, Stoltz *et al.* found different binding energies for the Pt atoms that adsorb above the surface Ti and O atoms of the STO support.[11] However, the electronic structure in such classical heterostructures has not previously been described, but would provide a clearer structure-function relationship.

For the surface of a crystal, XSW analysis is advantageous for determining atomic positions of a particular surface atomic species relative to the substrate lattice.[12–14] The collection of XSW excited fluorescence[10, 12, 15] or photoelectrons[16–20] from electronic core levels has previously been used for determining specific atomic adsorption sites.

The valence electronic structure dictates the chemical and physical characteristics of a supported catalyst that lead to the formation and breaking of chemical bonds during a reaction.[21] While density functional theory (DFT) can be used to calculate the projected density of states from particular atoms at particular sites within the substrate or at the interface,[22, 23] conventional valence band (VB) photoemission does not directly provide such structural discrimination. Previously, XSW excited VB spectroscopy has been used to study site-specific valence band properties for atoms within a bulk crystal.[24–27] Herein, we apply this XSW-VB method to the novel study of atoms on a surface. To increase sensitivity we chose a high-Z element, namely submonolayer Pt on STO (001). We find that XSW excited VB spectroscopy has site specific information from these surface atoms that compares well with DFT predictions.

The Pt submonolayer was grown on 1×1 TiO$_2$ terminated STO (001) surface by pulsed laser deposition (PLD). X-ray fluorescence with comparison to a calibrated standard was used to determine that the Pt coverage was ½ ML of a Pt (001) fcc layer (i.e.,

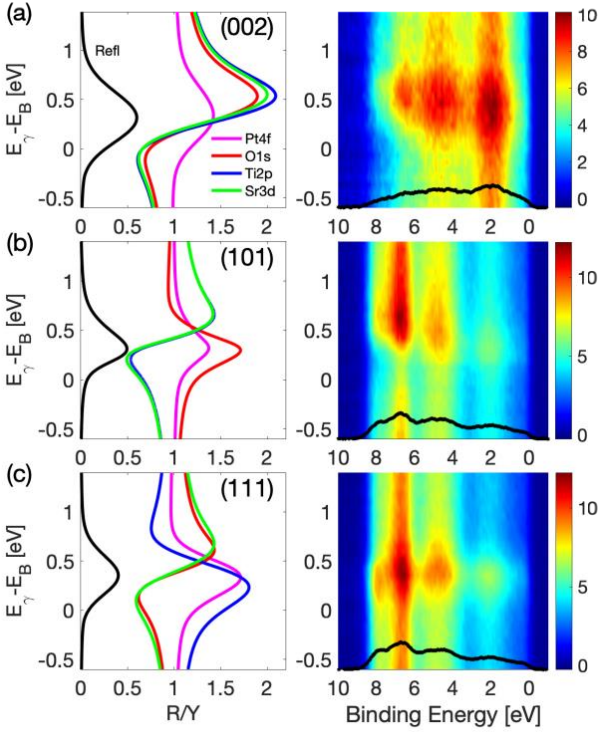

FIG. 1. (left-side) XSW excited Pt 4f, O 1s, Ti 2p and Sr 3d core-level yields and (right-side) valence band spectra collected while scanning the incident photon energy (E) through the (a) (002), (b) (101), and (c) (111) SrTiO3 Bragg reflections. ($E_B$ is the photon energy predicted by Braggs' law.) The black line at the bottom of each 2D VB spectrum is a plot of the VB spectrum collected at an off- Bragg energy. The 002 off-Bragg VB spectrum is different because it was collected at a shallower (more surface sensitive) emission angle.

TABEL I Measured coherent fraction $f_H$ and position $P_H$ values for Pt atoms for the 3 $hkl$ reflections. The model calculated values are determined from the global least-squares fit of the model described by Eq. 3 with values from Table 2.

| (hkl) | XSW Measured | | Model Calculated | |
|---|---|---|---|---|
| | $f_H$ | $P_H$ | $f_H$ | $P_H$ |
| (002) | 0.27 | 0.84 | 0.40 | 0.81 |
| (101) | 0.09 | 0.76 | 0.12 | 0.76 |
| (111) | 0.61 | 0.38 | 0.50 | 0.40 |

coverage = 6.5 Pt/nm$^2$). (See SI for details and Figs. S1-S3.) At the Diamond Light Source I09 station (Fig. S4-S5) back reflection XSW-XPS[13] was performed for three different Bragg reflections: (002), (101), (111) of the STO (001) crystal. Pt 4$f$, Ti 2$p$, Sr 3$d$ and O 1$s$ core level (CL) spectra (Fig. S6) and VB spectra were collected while scanning the incident photon energy (E) through each $hkl$ Bragg reflection. The XSW induced modulations in the CL and VB photoelectron yields are shown in Fig. 1.

For each **H** = $hkl$ Bragg reflection, Fig. 1 shows the CL data fitted yield curves based on dynamical diffraction theory:[13, 14, 28, 29]

$$Y(E_\gamma) = 1 + S_R R(E_\gamma) + 2|S_I|\sqrt{R(E_\gamma)} f_H \cos(\nu(E_\gamma) - 2\pi P_H + \psi). \quad (1)$$

(See Fig. S7-S10 for fitting details). $R(E_\gamma)$ and $\nu(E_\gamma)$ respectively denote the reflectivity and XSW phase. Parameters $f_H$ and $P_H$ are the normalized amplitude and phase of the **H**$^{th}$ Fourier component of the distribution of atoms being monitored by the spectrometer with elemental or even chemical-state specificity. $S_R$, $|S_I|$ and $\psi$, which are corrections for small non-dipolar effects in the photoelectric cross-section, are tabulated in Table S1 and described further in the SI. For our analysis we selected the origin of the STO cubic unit cell to be at the Sr site, which consequently is then the origin for each $P_H$ scale. The Eq. 1 fit-determined $f_H$ and $P_H$ values for Pt 4$f$ are listed in Table I. The fitted results for the substrate atoms, which are summarized in Table S2-S4, follow the ideal perovskite structure of SrTiO3. The summation of these XSW-XPS measured Fourier components and their symmetry equivalents generates a model-independent 3D map of the Pt atomic distribution:[10,15,30,31]

$$\rho(\mathbf{r}) = 1 + 2 \sum_{\substack{H \neq -H \\ H \neq 0}} f_H \cos[2\pi(P_H - \mathbf{H} \cdot \mathbf{r})]. \quad (2)$$

The 3D map is a projection of the Pt atomic distribution from the x-ray footprint (~0.1 mm$^2$) into a single STO cubic unit cell, which leads to fourfold and mirror symmetry along the c-axis of the STO (001) surface. The generated 3D map of the Pt atomic distribution referenced to the STO lattice is shown in Fig. 2(a), where it can be seen that the Pt atoms form an $fcc$-like lattice registered to the STO lattice with a vertical shift. Based on the same domain averaging symmetry conditions described above, there are three symmetry inequivalent Pt sites in this arrangement; labeled A, B, and C. Note C1 and C2 are symmetry equivalent.

To determine the occupation fraction, $\alpha$, of Pt in the A, B, and combined C sites, an $fcc$-like Pt model with two atomic layers is assumed based on the previously described XSW model-independent analysis portrayed in Fig. 2(a). Here the correlated Pt atoms are laterally constrained to occupy the A, B and C sites, with respective fractional unit cell displacements of $z_A$, $z_B$, and $z_C$ relative to the bulk-like SrO plane. The values z and are determined from a global least squares fit of a model to the measured complex Fourier components as follows:[10, 15]

$$F_H = f_H e^{2\pi i P_H} = [\alpha_A\, e^{2\pi i H \cdot r_A} + \alpha_B e^{2\pi i H \cdot r_B} + \alpha_C(e^{2\pi i H \cdot r_{C1}} + e^{2\pi i H \cdot r_{C2}})]. \quad (3)$$

Here, $r_A = (0, 0, z_A)$, $r_B = (½, ½, z_B)$, $r_{C1} = (½, 0, z_C)$, $r_{C2} = (0, ½, z_C)$ are unit cell fractional positions of the A, B and C sites, respectively. $d_H$ is the STO d-spacing. The XSW measured $f_H$ and $P_H$ values are taken from Table I. The best-fit results, assuming a Pt atomic spread of $\sigma = 0.2$ Å in the Debye-Waller factor, are summarized in Table II. The ratio of occupation fraction $\alpha_A : \alpha_B : \alpha_C = 1: 2: 3$ is notably different from the ratio of ideal *fcc* Pt lattice which would be 1: 1: 2; consistent with the C site being the bottom and therefore most populated layer. Summing all the occupation fractions at the three sites gives the fraction of correlated Pt atoms to be 0.61; meaning that 39% of the Pt are uncorrelated with the STO lattice.

TABLE II. Results from Eq. 3 model-dependent analysis for Pt.

| Site | z | a |
|---|---|---|
| A | 0.43 | 0.10 |
| B | 0.33 | 0.20 |
| C1+C2 | -0.06 | 0.31 |

The off-Bragg VB spectra shown at the bottoms of Fig. 1 (right-side) are shown with better detail in Fig. 2(b) to 2(d). These were collected on the low-photon-energy side of the Bragg peak at $E_\gamma - E_B = -1.1$ eV. To decompose these VB spectra into sub-spectra from the four atoms we use an approach similar to that used for an XSW-VB study of bulk STO (001)[26], but now extended to include surface atoms. While this site-specific VB description is a direct outcome from DFT as illustrated for our case in Fig. 2(e), it cannot be determined from standard VB spectroscopy. Referring to Fig. 1, we experimentally determine the site by observing which features in the VB fluctuate when the XSW antinode (or node) passes over a given bulk or surface atomic layer. This relies on Woicik's earlier finding that even though the valence electrons are spatially spread out from the atomic center, their photoelectric cross-section is highly localized at the atomic center.[32] To get around the ambiguity of multiple atom types positionally overlapping in a given *hkl* direction, we chose a set of *hkl* reflections that have differing combinations of overlapping atom types. This sorting strategy is demonstrated in Fig. 1 (left-side), which show the Eq. 1 fitted XSW modulations of Pt 4*f*, Sr 3*d*, Ti 2*p* and O 1*s* core level yields for the incident energy scans through the STO 002, 101 and 111 reflections. For the 002 all three bulk atom types overlap with each other, but not with the Pt atoms. This can be seen from the phases (or $P_{002}$ values) of the four modulations in Fig. 1(a) and can be understood from the ball-and-stick diagrams of Fig. 2(a), which show respectively the bulk atomic positions at $P_{002} = 0$ coinciding with the 002 planes and the inward shift of the Pt atoms from these same planes at $P_{002} = -0.16$. For the 101 XSW scan of Fig. 1(b) we see that the Ti and Sr bulk atoms are in phase with each other at $P_{101} = 0$ and out of phase with the bulk O atoms at $P_{101} = ½$. From Table I we see that $P_{101} = 0.76$ for the Pt atoms. We then use the 111 to put the Sr and O in phase with each other and out-of-phase with the Ti and Pt as seen in Fig. 1(c). The 2D plots on the right-side of Fig. 1 are the VB spectra from -1 to 10 eV in binding energy collected at each incident photon energy step of the XSW scan with a range matching that of the left-side CL XSW scans.

Each binding energy (BE) point in the VB spectra (Fig. 1 right-side) is dependent on E because the four atoms can have different $\Delta d/d$ diffraction plane positions. Due to the previously explained localized nature of the photoelectric effect, each atomic VB subspectrum has an XSW induced modulation that follows the CL yield for that atom as shown in Fig. 1 left-side. As detailed in the SI this relation is formalized with a matrix equation for each of the 251 binding energy ($BE_i$) points in each VB spectrum. The solutions to these overdetermined sets of equations for each of three *hkl* reflections leads to the decomposition of the off-Bragg VB spectrum into sub-spectra for each constituent element as shown in Fig. 2(b) to (d). For each element, the three VB sub-spectra from the three *hkl* reflections differ by a scale-factor that is due to differences in the experimental conditions, namely incident photon energy, incident intensity, sample-detector geometry, and photoelectric cross-sections. We were able to create an all-inclusive set of simultaneous equations by normalizing for these differences, which led to a solution with higher fidelity.

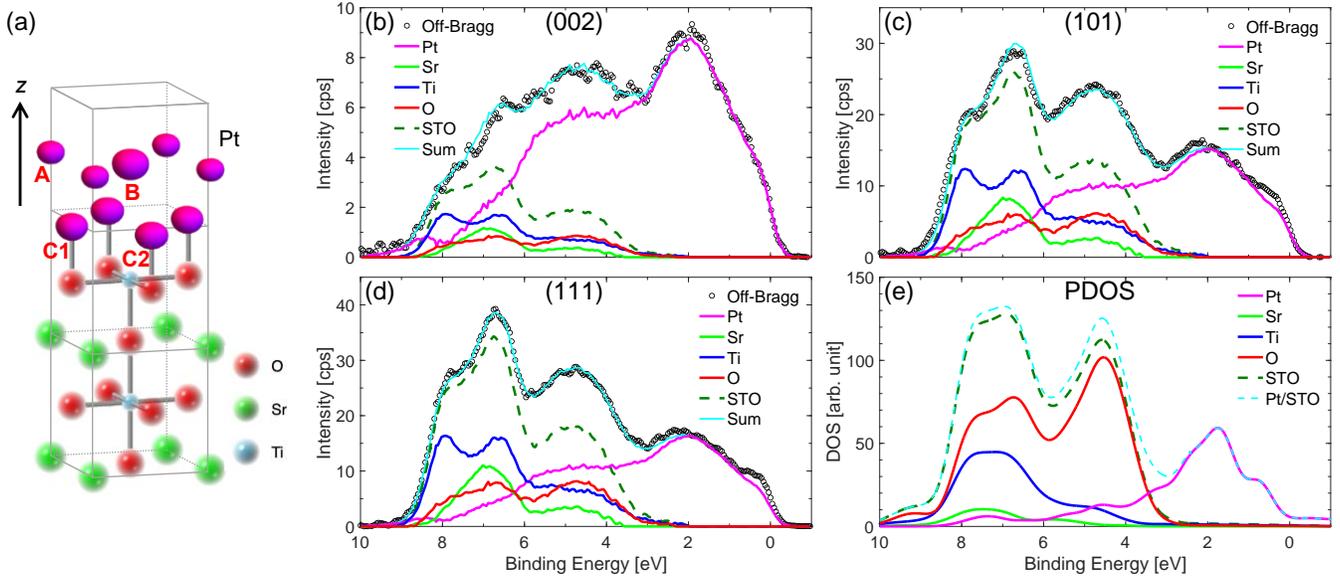

FIG. 2. (a) Based on Eq. 2, the model-independent XSW measured 3D Pt atomic distribution relative to the STO unit cell shown with a $TiO_2$ terminated STO (001) surface. The Pt contour plot is at 80% $\rho_{max}$. A, B, and C1(C2) refer to the three symmetry-inequivalent Pt sites. (b-d) The collected off-Bragg VB spectra and separated components from (b) (002), (c) (101), and (d) (111) $SrTiO_3$ Bragg reflections. Note that the (002) VB spectrum is more sensitive to Pt because of the lower photoelectron emission angle. (e) DFT calculated projected density of states (PDOS) of the Pt/STO valence band.

To compare with the XSW-XPS results, DFT calculations were carried out for the projected densities of states (PDOS) of each element using the Spanish Initiative for Electronic Simulations with Thousands of Atoms (SIESTA) code. The 3D model generated from the XSW-CL photoemission analysis was used to construct an atomistic super cell. The PDOS were computed using the conjugate-gradient (CG) algorithm, with the generalized gradient approximation (GGA) Perdew–Burke–Ernzerhof (PBE) functional and a broadening of 0.4 eV. The PDOS of each element as a summation of its valence electronic states is shown in Fig. 2(e).

In Fig. 3 we compare the DFT calculations to the experimental results by converting each orbital PDOS to photoelectron yield. (See SI for details.) Here the photoelectric effect cross section ratios are corrected because the solid-state VB is significantly underestimated by the tabulated atomic cross section.[25, 26] The Pt sub-spectrum in Fig. 3(a) has mainly Pt $5d$ states and $4f$ states located near the Fermi level. Note that the DFT calculated Pt sub-spectrum corresponds to correlated Pt as depicted in Fig. 2(a), whereas the measured Pt VB sub-spectrum has a 39% contribution from uncorrelated Pt atoms. Given this factor, the difference between calculated and experimental yields is reasonable. The calculated Sr yield in Fig. 3(b) reproduce the experimental yield through $5s$, polarized $5p$ and semicore $4p$ states. We note a peak shift close to 1 eV, Fig. 3(b), likely due to an inherent shift already present in the original SIESTA pseudopotential (free of semicore states). The calculated Ti yield in Fig. 3(c) includes $3d$ and $4s$ states and the Ti $3p$ semicore was introduced to create the features at 4 - 7 eV. The calculated O yield Fig. 3(d) perfectly reproduces the experimental yield with a majority of O $2p$ state. The features of Sr, Ti and O follow the previous XSW-VB results for bulk STO.[26] Compared with the valence band spectrum of bulk metallic Pt,[33] the Pt ML yield is missing a strong peak right at the Fermi edge, which could correspond to a reduction of electronic states due to the orbital hybridization of Pt $4f$ with O atoms. DFT was used to study this effect by comparing to a model with Pt atoms at C-sites only (Fig. 2(a)) located atop of O atoms without the top Pt layer (A and B sites). Here the Pt $4f$ states at the Fermi level is reduced to zero (Fig. S12). The round shoulder, rather than a peak, at the Fermi edge for the Pt sub-spectrum further confirms the Pt-Pt bond and Pt-O bond in the Pt model with two atomic layers. Future 001 XSW-XPS measurements of this system should discriminate between these two different Pt layers.

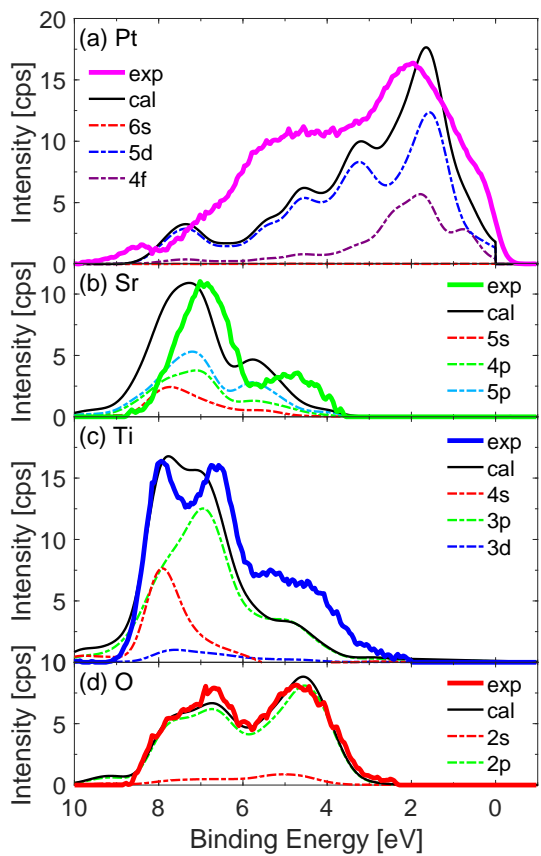

FIG. 3. Decomposed (a) Pt, (b) Sr, (c) Ti, and (d) O valence band spectra in (111) off-Bragg condition based on experimental data (exp) compared with DFT calculated PDOS (s, p, d or f states) converted to photoelectron yield (cal). Note that the sum of all four experimental sub-spectra is equivalent to the spectrum labeled "Sum" in Fig. 2(d).

In summary, the 3D atomic distribution of ½ ML Pt on the $SrTiO_3$ (001) surface was measured from the XSW excited core level Pt $4f$ yields for three $hkl$ Bragg reflections. This provided a model for a density functional theory calculation that was used to interpret the XSW site-specific valence band measurements. The XSW-XPS core level yields of Pt $4f$, Sr $3d$, Ti $2p$ and O $1s$ make it possible to separate out the Pt, Sr, Ti and O contributions to the valence-band spectrum. It is presently possible to make such measurements for monolayer coverages of high-Z surface atoms like Pt with 3$^{rd}$ generation synchrotron x-ray sources. The predicated $10^2$ gain in brightness from the next generation sources coming on line should make it possible to apply this method to a monolayer of surface atoms with much lower-Z.

**ACKNOWLEDGMENT**

This work was primarily supported by the Northwestern University (NU) Institute of Catalysis in Energy Processes which is funded by the U.S. Department of Energy, Office of Science, Office of Basic Energy Sciences under Award DE-FG02-03ER15457. We thank Diamond Light Source for access to beamline I09 (proposal number SI20076) where the data was collected. We thank Dave McCue and Pardeep Khakur for assistance at I09. We thank Jörg Zegenhagen and Ivan Varntanyants for a helpful discussion. We thank D. Bruce Buchholz for help in using the PLD Facility supported by the NU-MRSEC (NSF-DMR-1720139). Preliminary measurements were made at DND-CAT at the Advanced Photon Source (APS) supported by DuPont, Northwestern University (NU), and Dow Chemical. Argonne National Lab is sup- ported by DOE Grant No. DE-AC02-06CH11357. This research was supported in part by the computational resources and staff contributions provided by the Quest High Performance Computing Facility at NU, which is jointly supported by the Office of the Provost, the Office for Research, and NU Information Technology. This work made use of the XRD, PLD, and Keck-II facilities at NU supported by the MRSEC program of the (NSF DMR-1720139), Keck Foundation, State of Illinois, and the Soft and Hybrid Nanotechnology Experimental (SHyNE) Resource (NSF ECCS-1542205).

*bedzyk@northwestern.edu; Corresponding author

**Supplemental Material**

# Site-specific surface atom valence band structure via x-ray standing wave excited photoemission

Yanna Chen,[1] Leighton O. Jones,[2] Tien Lin Lee,[3] Anusheela Das,[1] Martín A. Mosquera,[4] Denis T. Keane,[1] George C. Schatz,[2] Michael J. Bedzyk[1,5,*]

[1] Department of Materials Science and Engineering, Northwestern University, Illinois 60208, USA
[2] Department of Chemistry, Northwestern University, Evanston, Illinois 60208, USA
[3] Diamond Light Source, Harwell Science and Innovation Campus, Didcot, OX11 0DE, UK
[4] Department of Chemistry and Biochemistry, Montana State University, Bozeman, Montana 59717, USA
[5] Department of Physics and Astronomy, Northwestern University, Illinois 60208, USA

*E-mail: bedzyk@northwestern.edu.

**Growth of Pt monolayer by PLD**

Strontium titanate ($SrTiO_3$) single crystals were oriented, cut (10×10×1 $mm^3$) with a miscut angle of ~0.1° and polished parallel to the (001) by MTI (USA). The $SrTiO_3$ (001) substrates were ultrasonicated for 30 minutes in deionized water (18 MΩ/cm) and subsequently etched in a buffered hydrofluoric acid for approximately 2 minutes. The substrates were then rinsed in deionized water and dried in $N_2$ gas. To produce atomically flat terraces terminated with titanium oxide, the substrates were loaded into a tube furnace with $O_2$ flow (~400 sccm) and annealed at 1050 °C for 5 hours. The surface topography of the sample before and after Pt deposition was measured by atomic force microscopy (AFM, Bruker ICON) in tapping mode. The STO surface after the $O_2$ anneal showed large flat terraces (Fig. S1(a)). Low-energy electron diffraction (LEED, OCI BDL800IR) of the STO surface showed a (1×1) pattern (Fig. S2(a)). These features are consistent with a low-defect $TiO_2$ terminated STO (001) surface. [1, 2] After the annealing process, the substrates were mounted in the pulsed laser deposition (PLD, PVD NanoPLD) chamber with a base pressure of 4.6×10$^{-8}$ Torr. The PLD process used 250 laser pulses on a Pt target with fluency of 1.24 $J/cm^2$ and a frequency of 5 Hz with a substrate temperature of 400 °C. The sample was then annealed inside the PLD chamber at a series of increasing temperatures of 500 °C for ½ h, at 700 °C for 1 h, and at 850 °C for 1 h. The heater power was then turned off and the sample was cooled down slowly to room temperature before removal from the PLD vacuum chamber. AFM of the Pt/STO surface (after the deposition and vacuum anneal) is shown in Fig. S1(b). LEED showed the same 1×1 pattern in Fig. S2(b) as prior to deposition.

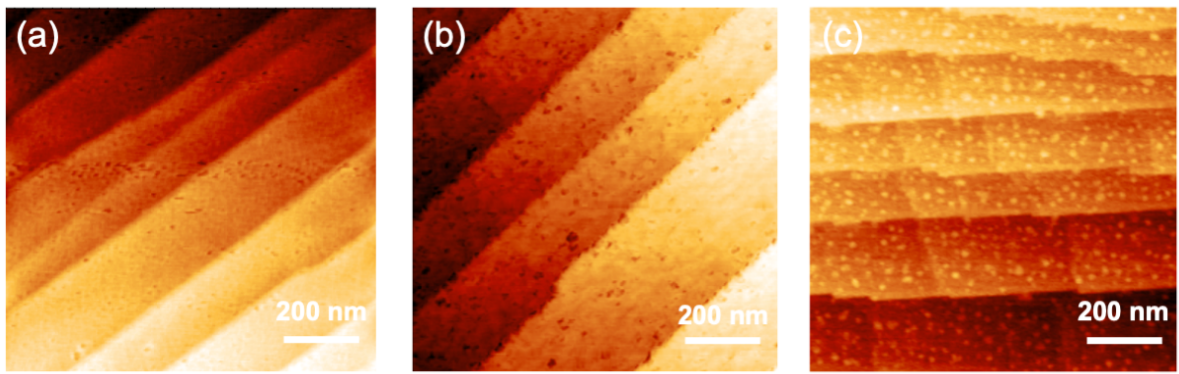

**Figure S1.** AFM images for $TiO_2$ terminated $SrTiO_3$ (001) surface (a) prior to and (b) after ½ ML and (c) 1.2 ML Pt deposition. The Pt islands was observed for 1.2 ML Pt.

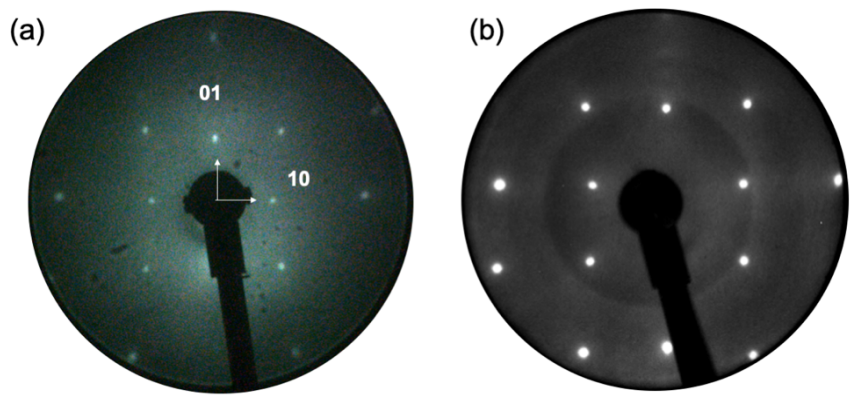

**Figure S2.** 1×1 LEED patterns from the $TiO_2$ terminated $SrTiO_3$ (001) surface (a) prior to and (b) after Pt deposition.

# X-ray fluorescence (XRF) Pt coverage measurement

XRF measurements were performed at the APS station 5BM-C of DND-CAT with an incident photon energy of 13.5 keV. A Vortex silicon drift diode detector was used to collect the x-ray fluorescence (XRF) spectra. An XRF spectrum of the Pt/SrTiO$_3$ sample is shown in Fig. S3(a). The deadtime corrected net counts per second of the Pt $L\alpha$ peak in the spectrum was determined by the fit of a Gaussian function on a linear background. The Pt coverage was determined by a side-by-side comparison to the Ga $K\alpha$ emission from a Ga implanted standard with a coverage of 12 Ga/nm$^2$, which was calibrated by Rutherford backscattering (RBS). The XRF spectrum of the Ga implanted standard measured at the same conditions is shown in Fig. S3(b). The Pt coverage was determined through comparison of area of Pt $L\alpha$ peak and that of Ga $K\alpha$ peak considering atomic subshell photoemission cross-sections, detector sensitivity and detector live time. At 13.5 keV the atomic XRF cross-section ratio of Ga $K\alpha$ to Pt $L\alpha$ is 0.769.

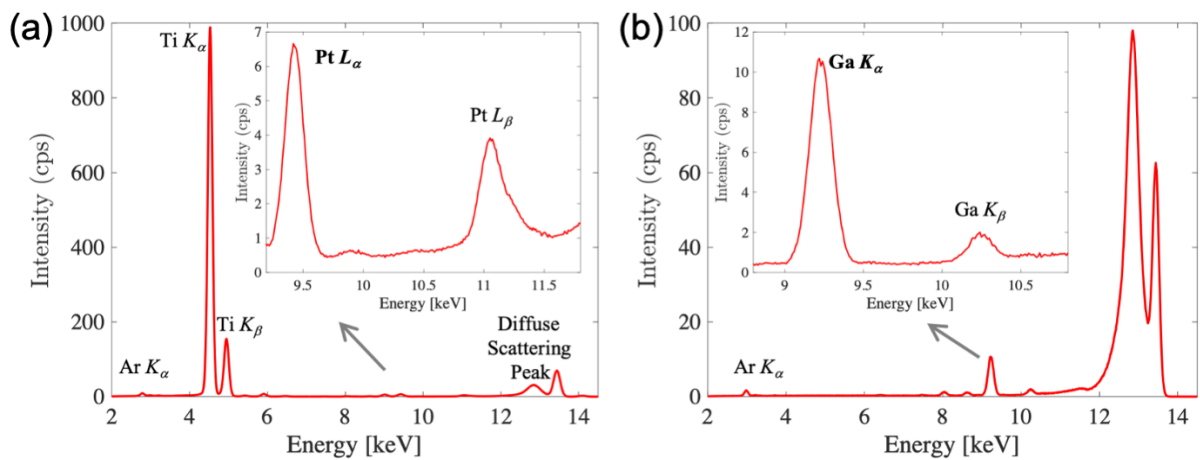

**Figure S3**. XRF spectra collected at an incident photon energy of 13.5 keV for the (a) Pt/SrTiO$_3$ (001) sample and (b) Ga implanted standard with 12 Ga/nm$^2$. From this side-by-side XRF analysis the Pt coverage was determined to be 6.5 Pt/nm$^2$. Based on a Pt $fcc$ (001) atomic plane with $a = 3.924$ Å this corresponds to ½ ML of Pt. The peak at 13.5 keV is from thermal diffuse scattered x-rays and the peak at 12.7 keV is from Compton scattered x-rays.

**Surface preparation and x-ray standing wave (XSW), x-ray photoelectron spectroscopy (XPS) and low-energy electron diffraction (LEED) measurements at Diamond I09**

The XSW measurements were carried out at the Diamond Light Source EH2 end-station of beamline I09 (Fig. S4). [3] The beamline can provide both soft (230 – 800 eV) and hard x-rays (2.1 – 18 keV) at the same spot on the sample from two separate undulators. Soft x-rays used a defocused beam size of 300×300 μm². Hard x-rays had a flux of $10^{13}$ photon/s and the beam size were focused to 120×120 μm². The UHV system in EH2 contains two surface preparation chambers (SPC-1 and SPC-2) and the analysis chamber. The Pt/SrTiO$_3$ (001) sample was loaded into SPC-2. SPC-2 is equipped with a hBN (ceramic) heater (accurate to ±10°). A thermal cracker cell (Oxford Applied Research TC50) was used to convert molecular O$_2$ gas into atomic O, which was used to efficiently clean the surface. The samples were heated to 300°C and an O$_2$ pressure of 5×10$^{-7}$ mbar was kept for 30 minutes. Then the sample was transferred to the analysis chamber, which has a higher resolution LEED. LEED patterns were acquired to verify the Pt/SrTiO$_3$ surface structure at the voltage of 154 V. Then XSW measurements were acquired using hard x-rays in a back-reflection geometry ($\theta_B$ = 88°) as shown in Fig. S5. The sample asymmetry factor $b_s = \sin(\alpha_0)/\sin(\alpha_h)$ depends on the reflection geometry, [4] which is shown in the right of Fig. S5 for specular and off-specular reflections. $b_s$ is equal to -1, -0.93 and -0.91 for (002), (101) and (111) reflections, respectively. The $b_s$ factor affects coherent fraction by up to 4% and coherent position by up to 2%. The Bragg diffracted beam in this near back-reflection geometry was measured by a CCD camera pointed at a Cu plate coated with YAG fluorescent powder. The incident x-ray beam energy ($E_\gamma$) was tuned through the first harmonic of the I09 undulator and the Si (111) double-crystal monochromator. The photon energy based on Braggs' law was $E_B$ = 3.18 keV, 2.25 keV and 2.76 keV for the SrTiO$_3$ (002), (101) and (111) reflections, respectively. The HAXPES analyzer is a VG Scienta EW4000 with a 70 frame/sec CCD camera. The XPS binding energy scale and resolution were calibrated using a Au Fermi edge. All the contributions (thermal broadening ~ 120 meV at room temperature, beamline resolution ~303 meV at 3 keV from the energy bandpass of Si (111) DCM, and analyzer resolution 0.25 eV for valence band and 0.05 for core level) combined quadratically had an energy resolution of 0.41 eV for valence-band spectra and 0.33 eV for core-level spectra at the photon energy $E_\gamma$ = 3.00 keV. The HAXPES data collection used a fixed mode of Angular 56. The HAXPES valence band spectra with a binding energy range of 30 eV were collected in 24 minutes scanning through a reflection, which included 41 data points.

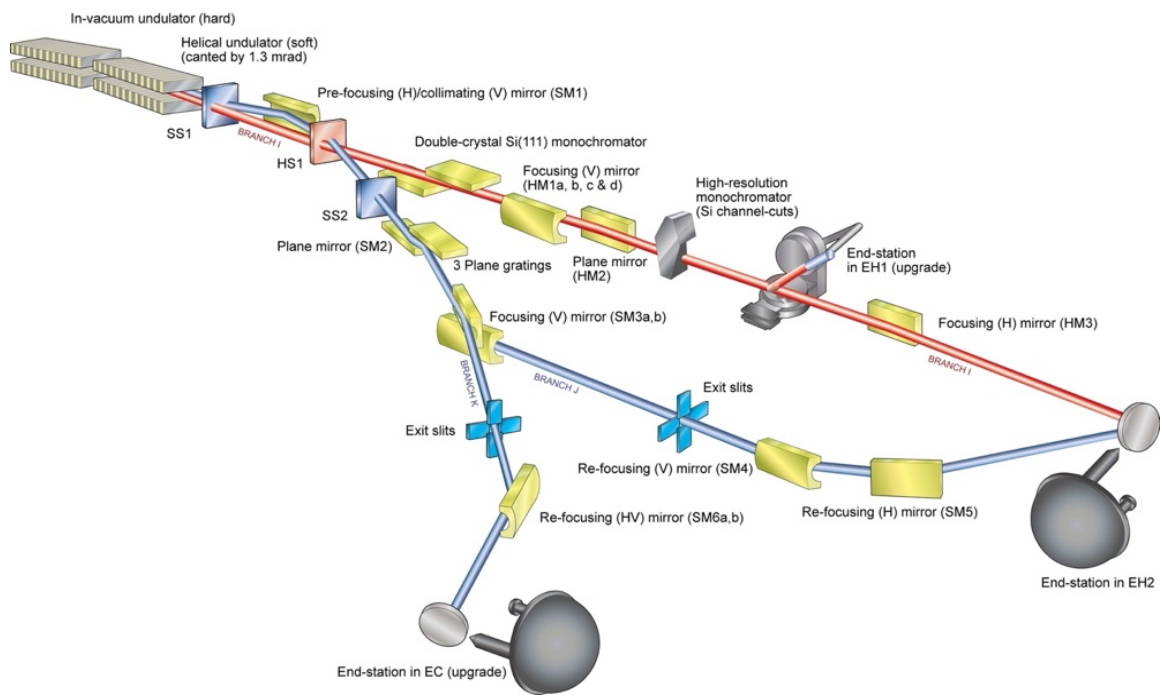

**Figure S4.** Diagram of the I09 beamline at the Diamond Light Source. Experiments were performed in the EH2 end-station using hard and soft x-rays from separate undulators.[3]

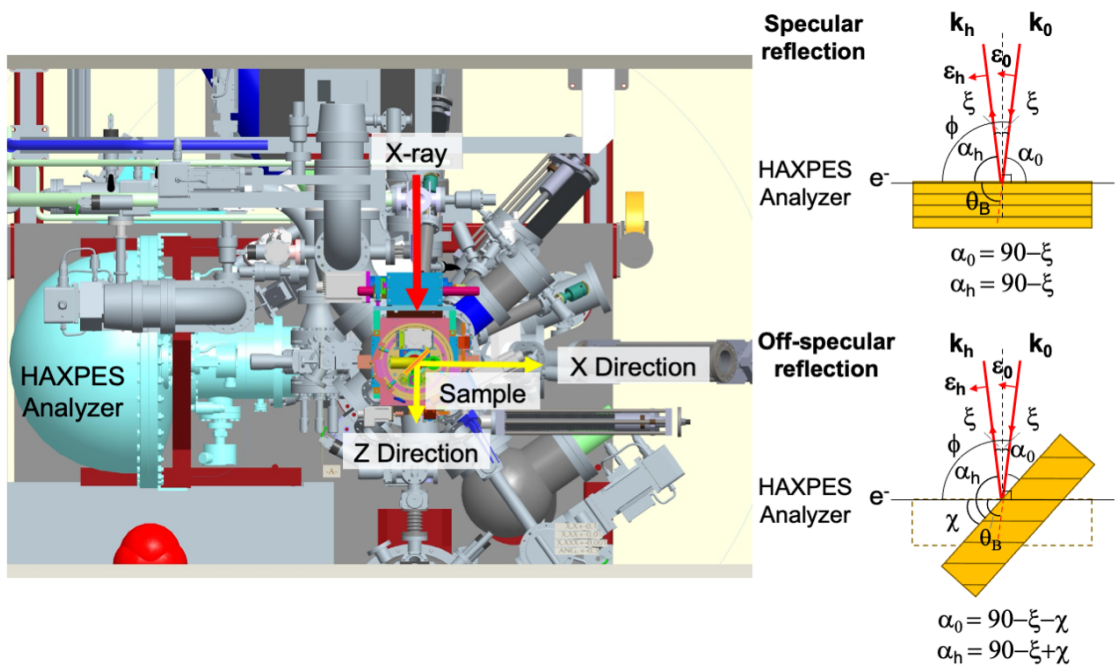

**Figure S5.** (Left-side) Back-reflection geometry in the UHV system at the EH2 end station (top view). The incident x-ray beam propagates in the Z-direction and is polarized in the X-direction. The cone axis of the HAXPES analyzer acceptance lens is in the X-direction. The sample surface is parallel with the vertical Y-axis and can be rotated about this Y-axis and about an azimuthal-axis perpendicular to the sample surface. (Right-side) Schematic top-view of the sample π-polarized x-ray scattering geometry for the *002* specular symmetric reflection (top) and off-specular asymmetric *101* and *111* reflections (bottom). The angle $2\xi = 180 - 2\theta_B$ is the deviation from ideal normal incidence diffraction geometry, where the Bragg angle $\theta_B$ is angle between the incident ($k_0$) or diffracted ($k_h$) x-ray beam and the *hkl* lattice plane. $\alpha_0$ is the angle between the incident beam and the sample surface. $\alpha_h$ is the angle between the diffracted beam and the sample surface. $\chi$ is the angle between the *hkl* reflection planes and the surface. $\phi$ is the angle between the photoelectron emission direction and the incident beam.

**Core level (CL) spectra**

Pt 4*f*, Ti 2*p*, Sr 3*d*, and O 1*s* core level (CL) spectra were collected with the Scienta EW4000 spectrometer in a snapshot mode with a maximum acceptance angle of the XPS detector of 58°. XPS data were analyzed using CasaXPS software.[5] To compenensate for charging effects, the energy scale for each photoelectron spectrum was shifted such that the adventitious carbon 1s occurred at a binding energy of $BE$ = 284.5 eV. This corresponded to subtracting 1.3, 3.4 and 2.2 eV from the binding energies of the as-collected spectra for the (002), (101) and (111) reflections, respectively. After this correction the binding energies for the Sr, Ti and O peaks all aligned with previously established values for SrTiO$_3$ single crystal.[6] Most importantly, this same $BE$ scale correction was applied to the VB spectral data shown in the manuscript and SI. The CL spectra were fitted after subtracting the Shirley background. Pt 4*f*, Ti 2*p* and Sr 3*d* are all doublet peaks. Because the two branches have a fixed ratio, we choose the higher branch. The integrated areas of the Pt 4$f_{7/2}$, Ti 2$p_{3/2}$, Sr 3$d_{5/2}$ and O 1*s* were used for each XSW excited photoelectron yield scan at each *hkl* reflectivition. The collected and the fitted spectra are shown in Fig. S6.

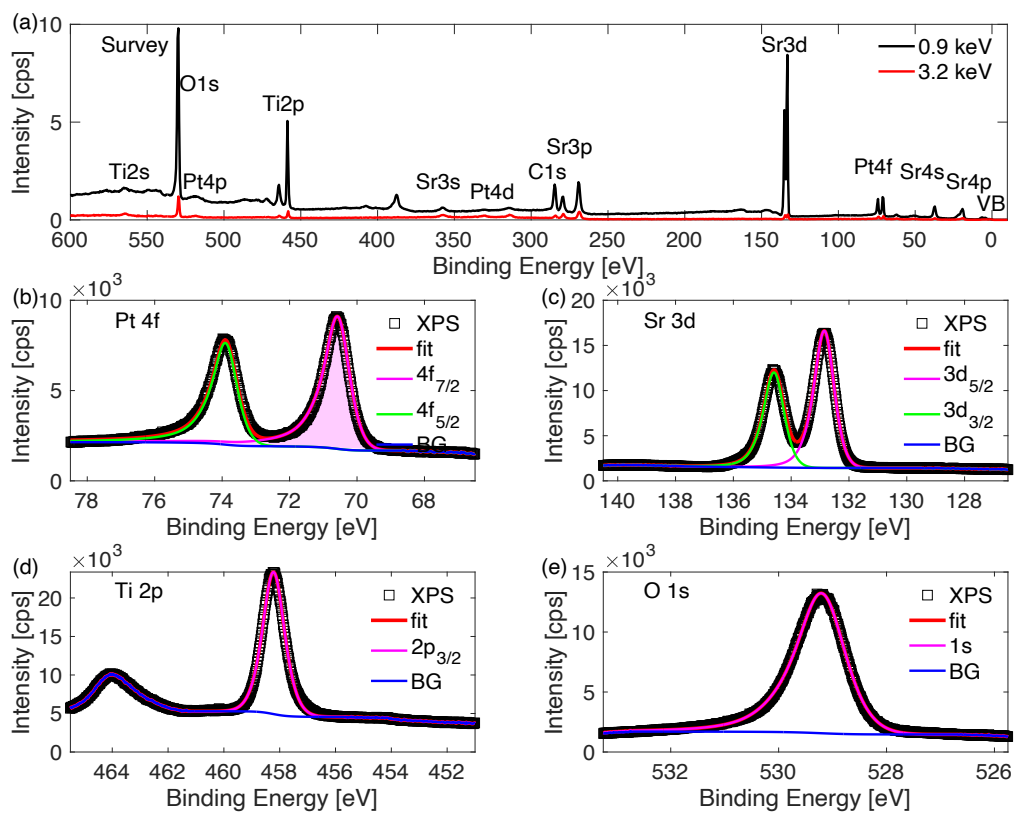

**Figure S6.** (a) Survey scan of Pt/STO collected at $E_\gamma$ = 0.9 keV and 3.2 keV. (b) Pt 4*f*, (c) Ti 2*p*, (d) Sr 3*d*, and (e) O 1*s* core level (CL) spectra from off-Bragg (002) reflection at 3.2 keV. Squares are the measured data. Red lines are fitted results. Magenta line and green line are the two fitted branches for the doublet peaks. Blue line is the Shirley background.

# X-ray standing wave analysis with non-dipolar corrections

The small non-dipolar corrections to the XSW photoemission yield of Eq. 1 are: [7]

$$S_R = \frac{1+Q'}{1-Q'}, \quad |S_I| = \frac{\sqrt{1+tan^2\psi}}{1-Q'}, \quad Q' = \frac{(\delta+\gamma cos^2\theta)sin\theta cos\varphi}{1+\beta P_2(cos\theta)}. \tag{S1}$$

Here $\beta$ is the dipole asymmetry parameter; $\delta$ and $\gamma$ are quadrupole parameters. These parameters can be obtained from Refs. [8, 9]. $P_2$ is second-order Legendre polynomial $P_2(cos\theta) = \frac{3cos^2\theta-1}{2}$. Angles $\theta$ and $\varphi$ are the spherical coordinates for the direction of the emitted photoelectron. The angle $\theta$ is the angle between the photoelectron emission direction and the x-ray polarization direction; $\varphi$ is the angle between the x-ray incident direction and the plane passing through the photoelectron emission direction and the x-ray polarization direction. For our experimental geometry the central axis of the acceptance cone for the photoelectron detector is aligned with the polarization direction. If the full cone is collected (as is the case for the off-specular reflections) the integration of $Q'$ over the full $2\pi$ radians in $\varphi$ leads to Q' = 0 and thus $S_R = |S_I| = 1$. For the specular (002) reflection the surface cuts out half of this cone, in which case we calculate $Q'$ by setting $\theta = 18°$ and $\varphi = 0$. The phase correction $\psi$ is related to $Q'$ and the phase shift difference $\Delta$ between the dipole and quadrupole matrix elements as:[7]

$$tan\psi = Q'tan\Delta. \tag{S2}$$

$\Delta = \delta_q - \delta_d$, where $\delta_q$ and $\delta_d$ are the elastic scattered phase shifts of transitions from initial state to quadrupole state or dipole state. Looking up the references [10, 11], $\psi$ changes the coherent position $P_H$ by less than 3%. Hence, $\psi$ is set as 0. All these parameters mentioned above are listed in Table S1.

**Table S1.** Photoemission nondipolar correction parameters used in Eq. (S1).

| | hkl | $E_\gamma$ [eV] | $KE_e$ [eV] | $\beta$ | $\delta$ | $\gamma$ | $\theta$ | $\phi$ | $Q'$ | $S_R$ | $|S_I|$ |
|---|---|---|---|---|---|---|---|---|---|---|---|
| Pt $4f_{7/2}$ | (002) | 3186 | 3115 | 0.957 | 0.158 | 0.708 | 18 | 0 | 0.136 | 1.314 | 1.157 |
| | (101) | 2255 | 2184 | 1.020 | 0.119 | 0.479 | 0 | 0 | 0 | 1 | 1 |
| | (111) | 2760 | 2689 | 0.986 | 0.141 | 0.611 | 0 | 0 | 0 | 1 | 1 |
| Sr $3d_{5/2}$ | (002) | 3186 | 3052 | 0.868 | 0.148 | 0.835 | 18 | 0 | 0.160 | 1.381 | 1.191 |
| | (101) | 2255 | 2121 | 0.999 | 0.109 | 0.670 | 0 | 0 | 0 | 1 | 1 |
| | (111) | 2760 | 2626 | 0.925 | 0.131 | 0.764 | 0 | 0 | 0 | 1 | 1 |
| Ti $2p_{3/2}$ | (002) | 3186 | 2727 | 1.095 | 0.078 | 0.836 | 18 | 0 | 0.133 | 1.307 | 1.154 |
| | (101) | 2255 | 1796 | 1.241 | 0.054 | 0.684 | 0 | 0 | 0 | 1 | 1 |
| | (111) | 2760 | 2301 | 1.158 | 0.067 | 0.774 | 0 | 0 | 0 | 1 | 1 |
| O $1s$ | (002) | 3186 | 2655 | 1.974 | 0.000 | 1.116 | 18 | 0 | 0.116 | 1.262 | 1.131 |
| | (101) | 2255 | 1724 | 1.983 | 0.000 | 0.880 | 0 | 0 | 0 | 1 | 1 |
| | (111) | 2760 | 2229 | 1.978 | 0.000 | 1.016 | 0 | 0 | 0 | 1 | 1 |

**Pt 4*f*, Sr 3*d*, Ti 2*p* and O 1*s* XSW-XPS results**

We measured Pt 4*f*, Sr 3*d*, Ti 2*p* and O 1*s* using XSW-XPS on the Pt/SrTiO$_3$ (001) surface. The normalized photoelectron yields and reflectivities for the three *hkl* SrTiO$_3$ Bragg reflections are shown in Figs. S7-S10. The fit determined coherent fractions and coherent positions for Sr, Ti and O are summarized in Tables S2-S4.

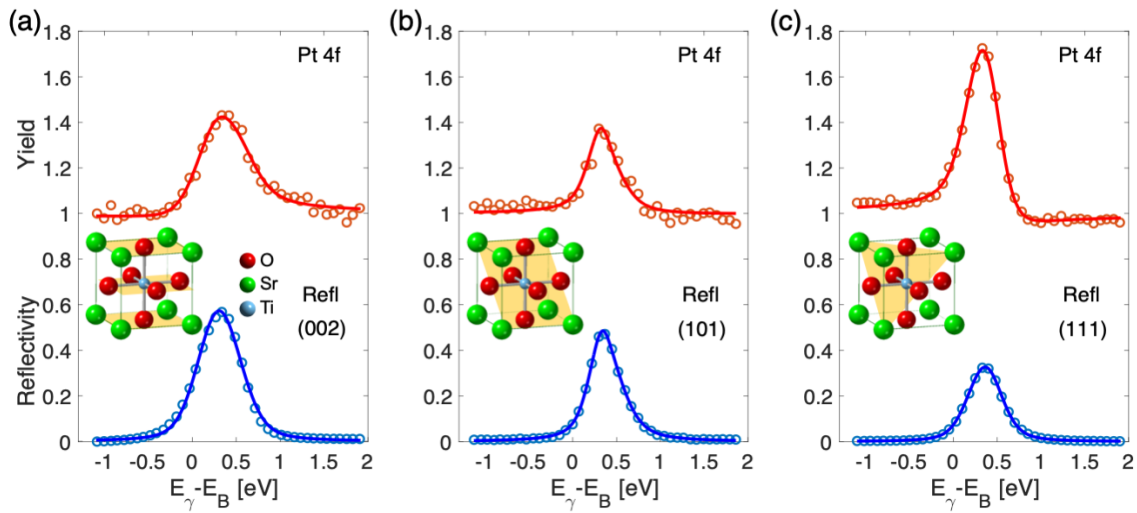

**Figure S7.** XSW-XPS measured Pt 4*f* normalized photoelectron yield (red circles) and reflectivity (blue circles) of the Pt/STO (001) surface for three STO Bragg reflections: (a) (002), (b) (101), and (c) (111) as a function of incident x-ray energy $E_\gamma$ relative to $E_B$, which is the Braggs' Law predicted x-ray energy for that *hkl* d-spacing at a fixed Bragg angle of $\theta_B = 88°$. The insets indicate the (*hkl*) planes in the STO unit cell at a position where $P_\mathbf{H} = 0$. The solid lines are fits to the data based on dynamical diffraction theory. The fit determined values for the coherent fraction and position are listed in Table 1.

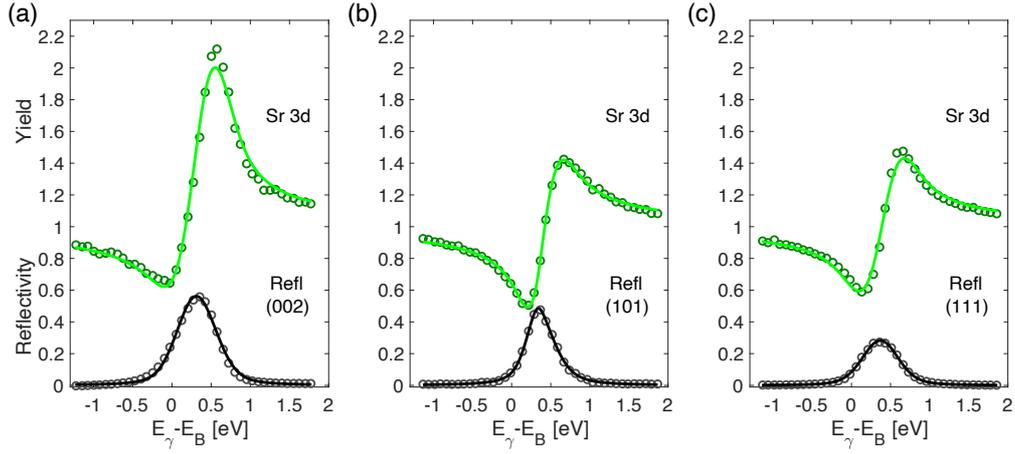

**Figure S8.** XSW-XPS measured Sr 3*d* normalized photoelectron yield (green circles) and reflectivity (black circles) of the Pt/SrTiO₃ (001) surface for three SrTiO₃ Bragg reflections: (a) (002), (b) (101), and (c) (111) as a function of $E_\gamma$ relative to $E_B$. The solid lines are fits to the data based on dynamical diffraction theory.

**Table S2.** Fit determined coherent fractions ($f_H$) and coherent positions ($P_H$) for Sr 3*d*.

| (hkl) | Experimental | | Ideal | |
|---|---|---|---|---|
| | $f_H$ | $P_H$ | $f_H$ | $P_H$ |
| (002) | 0.88 | 0.01 | 1 | 0 |
| (101) | 0.91 | -0.02 | 1 | 0 |
| (111) | 1.00 | 0.00 | 1 | 0 |

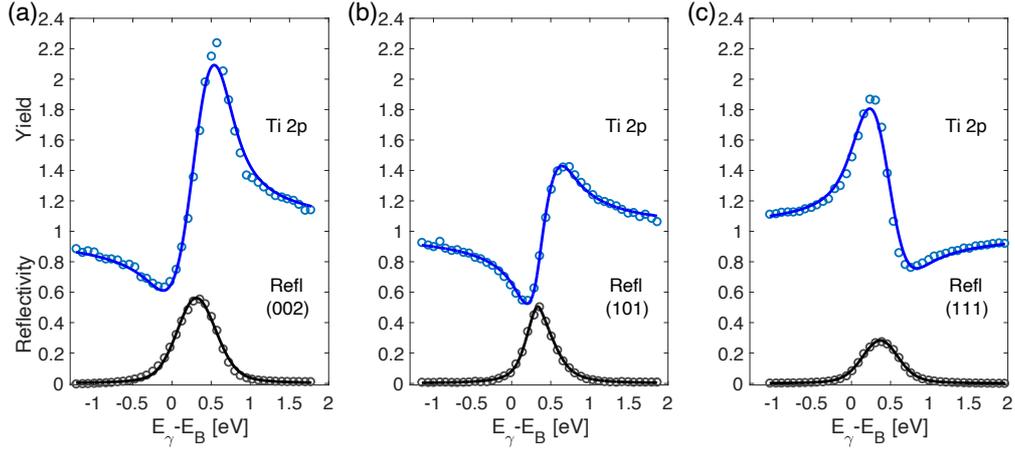

**Figure S9.** XSW-XPS measured Ti 2p normalized photoelectron yield (blue circles) and reflectivity (black circles) of the Pt/SrTiO$_3$ (001) surface for three SrTiO$_3$ Bragg reflections: (a) (002), (b) (101), and (c) (111). The solid lines are fits to the data based on dynamical diffraction theory.

**Table S3.** Fit determined coherent fractions ($f_H$) and coherent positions ($P_H$) for Ti 2p.

| (hkl) | Experimental | | Ideal | |
|---|---|---|---|---|
| | $f_H$ | $P_H$ | $f_H$ | $P_H$ |
| (002) | 0.96 | 0.03 | 1 | 0 |
| (101) | 0.85 | -0.01 | 1 | 0 |
| (111) | 1.00 | 0.49 | 1 | 0.5 |

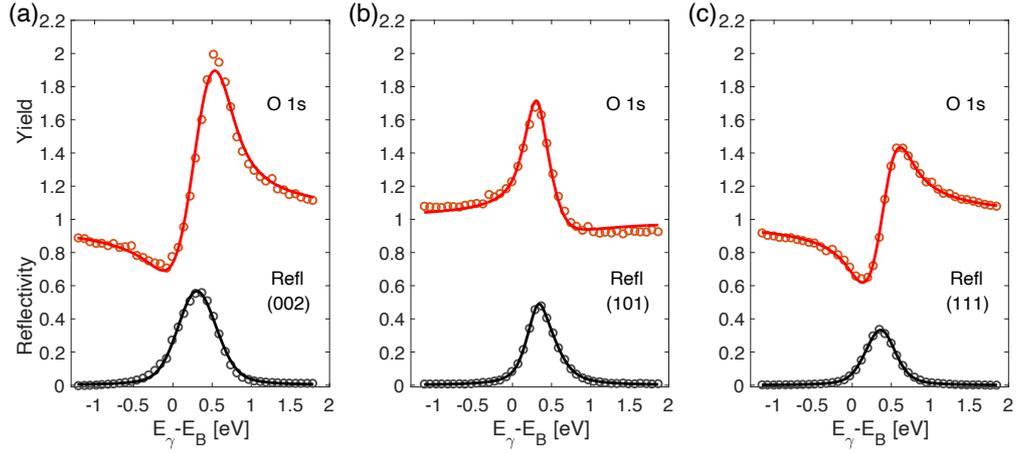

**Figure S10.** XSW-XPS measured O 1$s$ normalized photoelectron yield (red circles) and reflectivity (black circles) of the Pt/SrTiO$_3$ (001) surface for three SrTiO$_3$ Bragg reflections: (a) (002), (b) (101), and (c) (111). The solid lines are fits to the data based on dynamical diffraction theory.

**Table S4.** Fit determined coherent fractions ($f_H$) and coherent positions ($P_H$) for O 1$s$.

| (hkl) | Experimental | | Ideal | |
|---|---|---|---|---|
| | $f_H$ | $P_H$ | $f_H$ | $P_H$ |
| (002) | 0.79 | 0.01 | 1 | 0 |
| (101) | 0.33 | 0.51 | 0.33 | 0.5 |
| (111) | 1.00 | 0.00 | 1 | 0 |

**Decomposition of the valence band spectrum**

As depicted in Fig. S6, the four elemental contributions are easily separable in the core level (CL) photoemission spectrum. The same is not true for the valence band (VB) photoemission yield, since the four elemental contributions to the VB spectra overlap with each other as depicted in Fig. S11 open circles. To experimentally decompose the total VB spectrum into a sum of four sub-spectra we use the relation that the XSW induced modulation to the VB sub-spectra of a given element is proportional to the CL yield modulation of that element.[12] We now describe the matrix linear algebra solution for decomposing the VB spectrum into its elemental contributions based on this relationship.

As seen in Figs. S7 to S10, at each *hkl* Bragg reflection we measured XSW core-level yields for Pt 4*f*, Sr 3*d*, Ti 2*p*, and O 1*s* ($Y_{Pt}^{CL}$, $Y_{Sr}^{CL}$, $Y_{Ti}^{CL}$, $Y_{O}^{CL}$) at 41 incident photon energy steps $E_\gamma^j$ *(j = 1 to 41)* in the vicinity of the Bragg energy $E_B$. Eq. (S3) shows the corresponding 4 × 41 matrix *A* that contains all of these normalized core-level yield values. Correspondingly, there are 41 valence-band spectra in the binding energy range of -1 to 10 eV with 251 data points (*i* = 1 to 251). Each data point of the total valence-band yield $Y^{VB}(BE_i, E_\gamma^j)$ at the *i*th binding energy ($BE_i$) is composed of sub VB yields from Pt, Sr, Ti and O atoms, labelled as $Y_{Sr}^{VB}(BE_i)$, $Y_{Sr}^{VB}(BE_i)$, $Y_{Sr}^{VB}(BE_i)$, $Y_{Sr}^{VB}(BE_i)$, which are unknown forming matrix $B_i$ in Eq. (S4). At a fixed $BE_i$, the total valence-band yield, $Y^{VB}(BE_i, E_\gamma^j)$, is modulated by the XSW depending on the incident photon energy $E_\gamma^j$ *( j = 1 to 41)* forming matrix $C_i$ (Eq. (S5)). $C_i$ represents a (fixed $BE_i$) line cut in the 2D VB spectra of Fig. 1 right-side. Solving the overdetermined set of simultaneous equations represented by Eq. (S6), produces the four off-Bragg valence-band sub-spectra $Y_{Pt}^{VB}(BE_i)$, $Y_{Sr}^{VB}(BE_i)$, $Y_{Ti}^{VB}(BE_i)$, and $Y_{O}^{VB}(BE_i)$ from matrix $B_i$ with *i* = 1 to 251 .

$$A = \begin{vmatrix} Y_{Sr}^{CL}(E_\gamma^1) & Y_{Ti}^{CL}(E_\gamma^1) & Y_O^{CL}(E_\gamma^1) & Y_{Pt}^{CL}(E_\gamma^1) \\ Y_{Sr}^{CL}(E_\gamma^2) & Y_{Ti}^{CL}(E_\gamma^2) & Y_O^{CL}(E_\gamma^2) & Y_{Pt}^{CL}(E_\gamma^2) \\ \vdots & \vdots & \vdots & \vdots \\ Y_{Sr}^{CL}(E_\gamma^{41}) & Y_{Ti}^{CL}(E_\gamma^{41}) & Y_O^{CL}(E_\gamma^{41}) & Y_{Pt}^{CL}(E_\gamma^{41}) \end{vmatrix} \quad (S3)$$

$$B_i = \begin{vmatrix} Y_{Sr}^{VB}(BE_i) \\ Y_{Ti}^{VB}(BE_i) \\ Y_O^{VB}(BE_i) \\ Y_{Pt}^{VB}(BE_i) \end{vmatrix} \quad (S4)$$

$$C_i = \begin{vmatrix} Y^{VB}(BE_i, E_\gamma^1) \\ Y^{VB}(BE_i, E_\gamma^2) \\ \vdots \\ Y^{VB}(BE_i, E_\gamma^{41}) \end{vmatrix} \quad (S5)$$

$$A B_i = C_i \quad (S6)$$

In the first stage of our analysis we decompose off-Bragg total VB spectrum of each reflection into sub-spectra from spatially nonoverlapping atoms. For the (002) reflection, the normalized XSW modulations for the Sr, Ti and O overlap with each other, but not with that of Pt. Hence in matrix A,

$$Y_{Sr}^{CL}(E_\gamma^j) = Y_{Ti}^{CL}(E_\gamma^j) = Y_O^{CL}(E_\gamma^j) = Y_{STO}^{CL}(E_\gamma^j), \quad j = 1 \text{ to } 41. \quad (S7)$$

Therefore, for the 002 case the $A$ and $B_i$ matrices can be reduced to:

$$A = \begin{vmatrix} Y_{STO}^{CL}(E_\gamma^1) & Y_{Pt}^{CL}(E_\gamma^1) \\ Y_{STO}^{CL}(E_\gamma^2) & Y_{Pt}^{CL}(E_\gamma^2) \\ \vdots & \vdots \\ Y_{STO}^{CL}(E_\gamma^{41}) & Y_{Pt}^{CL}(E_\gamma^{41}) \end{vmatrix}, \quad B_i = \begin{vmatrix} Y_{STO}^{VB}(BE_i) \\ Y_{Pt}^{VB}(BE_i) \end{vmatrix}. \quad (S8)$$

The Eq. (S6) overdetermined set of 41 equations for each $BE_i$ are solved using a *Python* program with the *scipy* package. Values for matrix $B_i$ are obtained through a least square method using the wrapper *scipy.optimize.nnls* with the constraint that the matrix elements be nonnegative. This allows us to decompose the off-Bragg 002 VB spectrum into two VB sub-spectra; one for Pt and one for STO as shown in Fig. S11(a).

For the (101) reflection, the normalized XSW modulation for Sr and Ti overlap with each other, but not with that of O or Pt. Hence,

$$Y_{Sr}^{CL}(E_\gamma^j) = Y_{Ti}^{CL}(E_\gamma^j), \quad j = 1 \text{ to } 41. \tag{S9}$$

The three VB sub-spectra, Pt, O and SrTi, are seperated out and shown in Fig. S11(b). Since the Pt and O sub-spectra are not separable due to similar CL XSW modulation phases as seen in Fig. 1(b), we merge these two sub-spectra at this stage.

For the (111) reflection, XSW modulations for Sr and O overlap. Hence,

$$Y_{Sr}^{CL}(E_\gamma^j) = Y_O^{CL}(E_\gamma^j), \quad j = 1 \text{ to } 41. \tag{S10}$$

The three resolved VB sub-spectra for Pt, Ti and SrO are shown in Fig. S11(c).

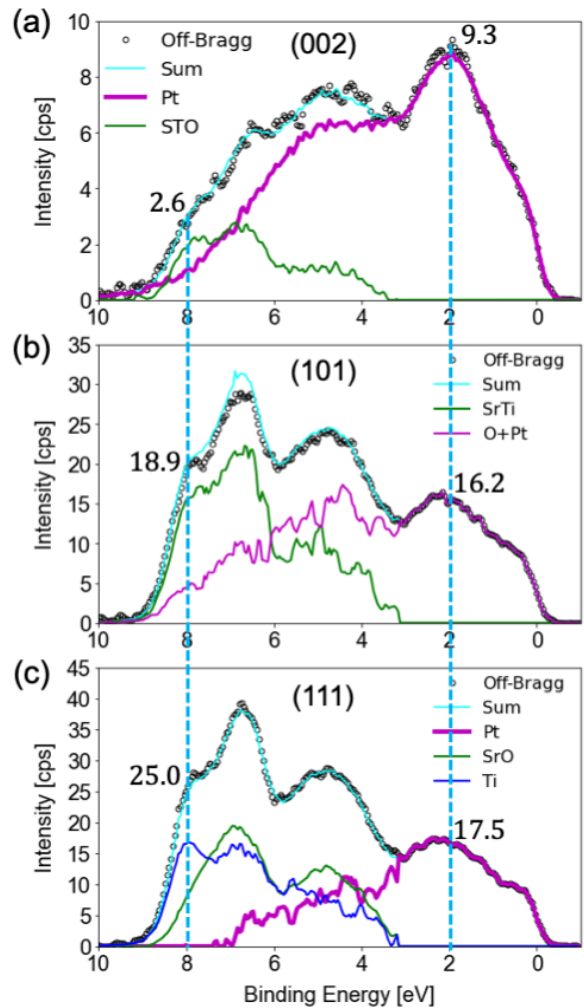

**Figure S11.** Valence-band spectral components that are contributed from combinations of Pt, Sr, Ti and O atoms. These are derived from the core-level and valence band yields obtained from the (a) (002), (b) (101) and (c) (111) reflections. The open circles are the off-Bragg VB spectra. The blue dash lines mark the measured VB yield at $BE = 8$ or $2$ eV.

In the second stage of the analysis, we complete the decomposition of the VB spectrum into sub-spectra from each element. As described above and shown in Fig. S11, we are able to directly separate out the Pt VB sub-spectrum from the (002) and from the (111) data. The (111) result (Fig. S11(c)) also isolates the Ti VB sub-spectrum. The Sr and O VB sub-spectra are not so simply separated from one individual $hkl$ data set due to an XSW modulation overlap with one of the other three elements as described by Eq. (S7) for the (002), Eq. (S9) for the (101) and Eq. (S10) for (111). Therefore, we will combine the results from the three $hkl$ XSW-XPS data sets into a single matrix equation to separate out the Sr and O VB sub-spectra. This treatment will also better refine the Pt and Ti sub-spectra. This requires a normalization of the three spectra, since each of the three different spectra was collected at a different incident photon energy ($E_\gamma$), incident photon intensity ($I_0$), sample - detector geometry, and with different photoelectric cross-sections ($\sigma$). For the STO substrate atoms, we will assume that the fractional atomic contributions to the off-Bragg STO VB spectrum do not change significantly, since the photoelectric cross-sections for each of the STO substrate atoms reduces uniformly as $E_\gamma$ increases from 2255 eV for the (101) to 2760 eV for the (111) to 3186 eV for the (002) as listed in Table S5. Therefore, we can determine the normalization factor adjustment to each individual atomic off-Bragg VB sub-spectra from the STO substrate by determining the overall normalization factor change for the STO contribution to the off-Bragg VB spectrum from each $hkl$. Since the fractional atomic contributions to the three off-Bragg STO VB spectra are assumed to be invariant and the feature at $BE$ = 8 eV is primarily due to STO and the features at $BE$ = 0 - 3 eV are due to Pt (Fig. S11 (a) and (c)), the counts per second at $BE$ = 8 eV and 2 eV can be used to calculate the normalization factors for the STO and Pt VB sub-spectrum, respectively. For example, the normalization factor $\eta_{STO}^{111}$ for comparing the (111) to (002) STO VB sub-spectra can be calculated as the ratio of the counts per second at $BE$ = 8 eV in Fig. S11.

$$\eta_{STO}^{111} = \frac{Y_{STO111}^{VB} at\ BE=8\ eV}{Y_{STO002}^{VB} at\ BE=8\ eV} = \frac{25.0\ cps}{2.6\ cps} = 9.47. \tag{S11a}$$

Similarly, the other normalization factors are:

$$\eta_{STO}^{101} = 7.16, \qquad \eta_{Pt}^{111} = 1.87, \qquad \eta_{Pt}^{101} = 1.74. \tag{S11b}$$

Using this normalization scheme, where for example $Y_{Sr101}^{VB}(BE_i) = \eta_{STO}^{101} Y_{Sr002}^{VB}(BE_i)$, we can rewrite Eq. (S4) as:

$$B_i^{002} = \begin{vmatrix} Y_{Sr002}^{VB}(BE_i) \\ Y_{Ti002}^{VB}(BE_i) \\ Y_{O002}^{VB}(BE_i) \\ Y_{Pt002}^{VB}(BE_i) \end{vmatrix} \text{ for (002) results,} \tag{S12}$$

$$B_i^{101} = \begin{vmatrix} \eta_{STO}^{101} Y_{Sr002}^{VB}(BE_i) \\ \eta_{STO}^{101} Y_{Ti002}^{VB}(BE_i) \\ \eta_{STO}^{101} Y_{O002}^{VB}(BE_i) \\ \eta_{Pt}^{101} Y_{Pt002}^{VB}(BE_i) \end{vmatrix} \text{ for (101) results,} \tag{S13}$$

$$B_i^{111} = \begin{vmatrix} \eta_{STO}^{111} Y_{Sr002}^{VB}(BE_i) \\ \eta_{STO}^{111} Y_{Ti002}^{VB}(BE_i) \\ \eta_{STO}^{111} Y_{O002}^{VB}(BE_i) \\ \eta_{Pt}^{111} Y_{Pt002}^{VB}(BE_i) \end{vmatrix} \text{ for (111) results.} \tag{S14}$$

Where matrix equation Eq. (S6) was for each individual *hkl*, this all-inclusive normalization scheme now allows us to combine all 3 individual matrix equations into one by moving the $\eta$ normalization factors into matrix *A*. This defines a new matrix *A'*.

$$A' = \begin{vmatrix} Y_{Sr002}^{CL}(E_\gamma^1) & Y_{Ti002}^{CL}(E_\gamma^1) & Y_{O002}^{CL}(E_\gamma^1) & Y_{Pt002}^{CL}(E_\gamma^1) \\ \vdots & \vdots & \vdots & \vdots \\ Y_{Sr002}^{CL}(E_\gamma^{41}) & Y_{Ti002}^{CL}(E_\gamma^{41}) & Y_{O002}^{CL}(E_\gamma^{41}) & Y_{Pt002}^{CL}(E_\gamma^{41}) \\ \eta_{STO}^{101} Y_{Sr101}^{CL}(E_\gamma^1) & \eta_{STO}^{101} Y_{Ti101}^{CL}(E_\gamma^1) & \eta_{STO}^{101} Y_{O101}^{CL}(E_\gamma^1) & \eta_{Pt}^{101} Y_{Pt101}^{CL}(E_\gamma^1) \\ \vdots & \vdots & \vdots & \vdots \\ \eta_{STO}^{101} Y_{Sr101}^{CL}(E_\gamma^{41}) & \eta_{STO}^{101} Y_{Ti101}^{CL}(E_\gamma^{41}) & \eta_{STO}^{101} Y_{O101}^{CL}(E_\gamma^{41}) & \eta_{Pt}^{101} Y_{Pt101}^{CL}(E_\gamma^{41}) \\ \eta_{STO}^{111} Y_{Sr111}^{CL}(E_\gamma^1) & \eta_{STO}^{111} Y_{Ti111}^{CL}(E_\gamma^1) & \eta_{STO}^{111} Y_{O111}^{CL}(E_\gamma^1) & \eta_{Pt}^{111} Y_{Pt111}^{CL}(E_\gamma^1) \\ \vdots & \vdots & \vdots & \vdots \\ \eta_{STO}^{111} Y_{Sr111}^{CL}(E_\gamma^{41}) & \eta_{STO}^{111} Y_{Ti111}^{CL}(E_\gamma^{41}) & \eta_{STO}^{111} Y_{O111}^{CL}(E_\gamma^{41}) & \eta_{Pt}^{111} Y_{Pt111}^{CL}(E_\gamma^{41}) \end{vmatrix}. \tag{S15}$$

Correspondingly, the $C_i$ matrix can be updated to $C_i'$ as:

$$C'_i = \begin{vmatrix} Y^{VB}_{002}(BE_i, E^1_\gamma) \\ \vdots \\ Y^{VB}_{002}(BE_i, E^{41}_\gamma) \\ Y^{VB}_{101}(BE_i, E^1_\gamma) \\ \vdots \\ Y^{VB}_{101}(BE_i, E^{41}_\gamma) \\ Y^{VB}_{111}(BE_i, E^1_\gamma) \\ \vdots \\ Y^{VB}_{111}(BE_i, E^{41}_\gamma) \end{vmatrix}. \tag{S16}$$

And finally, the all-inclusive matrix equation is:

$$A' B^{002}_i = C'_i. \tag{S17}$$

The resulting solutions for the four off-Bragg VB sub-spectra from the (002) reflection in matrix $B^{002}_i$ are shown in Fig. 3(a). The off-Bragg VB sub-spectra from the (101) and (111) reflections, shown in Fig. 3(b) and 3(c), are obtained through Eq. (S13) and Eq. (S14), respectively.

**Table S5.** Photoelectric cross-sections σ [Å²] for atomic core levels just below the valence band at the three different energies used for the three different *hkl* back-reflection XSW-XPS measurements. [13]

| *hkl* | $E_\gamma$ [eV] | $\sigma_{Sr}$ Sr 4$p_{3/2}$ BE = 23 eV | $\sigma_{Ti}$ Ti 3$p_{3/2}$ BE = 33 eV | $\sigma_O$ O 2s BE = 42 eV | $\sigma_{Sr} : \sigma_{Ti} : \sigma_O$ | $\sigma_{Pt}$ Pt 5$p_{3/2}$ BE = 52 eV |
|---|---|---|---|---|---|---|
| 101 | 2255 | 2.615×10⁻⁵ | 2.114×10⁻⁵ | 5.897×10⁻⁶ | 4.43 : 3.58 : 1 | 6.800×10⁻⁵ |
| 111 | 2760 | 1.629×10⁻⁵ | 1.177×10⁻⁵ | 3.413×10⁻⁶ | 4.77 : 3:45 : 1 | 4.759×10⁻⁵ |
| 002 | 3186 | 1.164×10⁻⁵ | 7.787×10⁻⁶ | 2.342×10⁻⁶ | 4.97 : 3.32 : 1 | 3.700×10⁻⁵ |

**DFT calculation**

The 3D atomic distribution of Pt on the SrTiO$_3$ surface has three sites, A (0, 0, $z_A$), B (½, ½, $z_B$) and C (½, 0, $z_C$) or (0, ½, $z_C$). Because the occupation fraction of the A site is much less than the B and C sites, we omitted the A site from the modeling and set up two models, one with Pt atoms at the B and C sites (Fig. S12 (a)) and the other with Pt atoms at the C sites only (Fig. S12 (b)). The pseudopotentials for Pt, Ti, and Sr, were regenerated using the ATOM 4.2.7 program (part of SIESTA) to include the semicore shells: Pt 4$f$, Ti 3$p$, and Sr 4$p$. The pseudopotential cutoff parameters of the Pt 6$s$, 6$p$ (empty), 5$d$, and 4$f$ shells are 2.38, 2.50, 2.38, and 1.98 bohr, respectively; for Ti the 4$s$, 3$p$, 3$d$, and 4$f$ (empty) cutoffs are 2.71, 2.45, 2.58, and 2.45 bohr, respectively; the Sr 5$s$, 4$p$, 4$d$ (empty), and 4$f$ (empty) cutoff radii are 3.58, 3.36, 3.20, and 3.58 bohr, respectively. The SIESTA numerical basis set for these atoms is DZP, except for the semicore shells, which are described with single zeta, SZ orbitals. Mesh cutoffs and filter cutoffs of 200 and 150 Ry were used, respectively. The energy tolerance was $1.0\times10^{-4}$ and the density matrix tolerance was $1.0\times10^{-5}$. Fermi–Dirac smearing was applied at an electronic temperature of 300 K, with a gamma-centered Monkhorst–Pack $k$-grid scheme of 16×16×1. The calculated density of states (DOS) of Pt is shifted relative to that of SrTiO$_3$ by 1.75 eV because of the interface barrier. [14] From the projected DOS (PDOS), we find that the Pt contribution to the valence band mainly arises from the 5$d$ and 4$f$ states. In Fig. S12 (a), the Pt 4$f$ state has two features at 1.8 eV and at 0.8 eV near the Fermi level. In Fig. S12 (b), the Pt 4$f$ state has only the feature at 1.8 eV. That means there are two kinds of Pt atoms in the monolayer. Pt atoms at C sites interacted with O atoms from the TiO$_2$ terminated layer forming a Pt-O bond, which contributes to the states at 1.8 eV. Other Pt atoms have Pt-Pt bonds contributing to the states near the Fermi level.

Converting the calculated PDOS to the elemental VB sub-spectrum, we consider photoelectric cross-sections, electron numbers, and introduce a correction factor to tune the deviation of the atomic and ionic photoelectric cross-section. The photoelectron yield, $Y(BE_i)$, converted from PDOS, $\rho_l(BE_i)$, is the summation of shells (e.g. $s, p, d, f$ shells with angular momentum quantum number $l$ = 0, 1, 2, 3) of each element:[15]

$$Y(BE_i) = \sum_{l=0,1,2\cdots} \kappa_l \sigma_l (1+\beta_i) \frac{N_l}{2J_l+1} \rho_l(BE_i). \qquad (S18)$$

Here, $\sigma$ is the total cross-section of the subshell; $\beta$ is a dipole parameter; $N$ is the electron number; $J$ is the quantum number of the total angular momentum of the electron. For the $l$ shell, the total photoelectric cross-section is $\sigma' = \sigma_l(1+\beta_l)\frac{N_l}{2J_l+1}$. $\kappa$ is correction factor for $\sigma'$. All parameters are listed in Table S6. Since solid state photoelectric cross-sections deviate significantly from atomic cross-sections, the relative ratio of the orbital cross-sections is corrected to agree with the experimental spectra. The resulting corrected yields based on the theoretical PDOS are shown in Fig. 4.

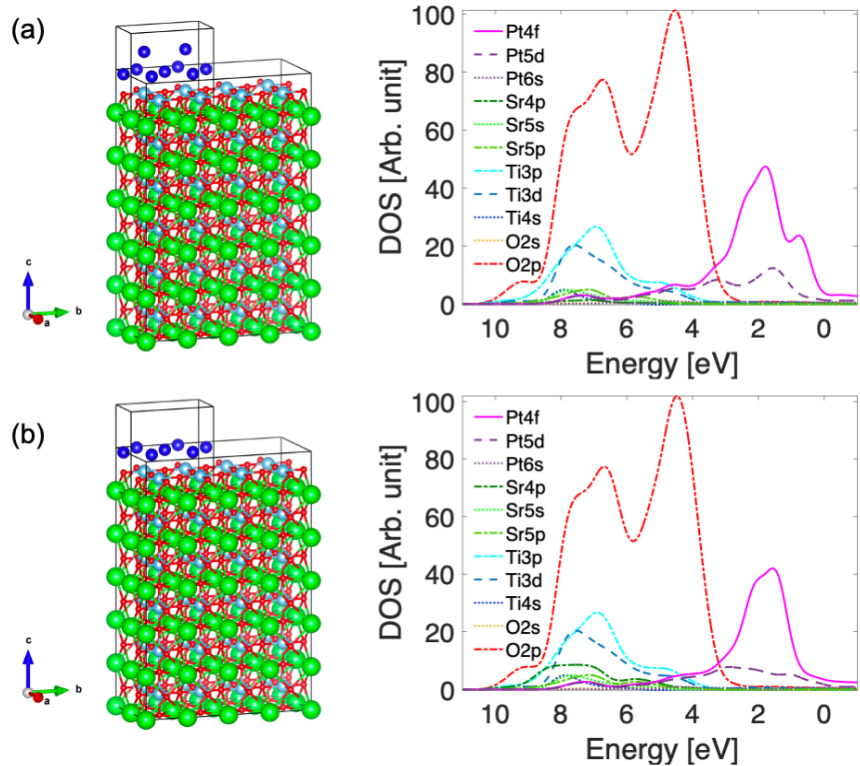

**Figure S12.** Two models of Pt/STO with respective DFT-calculated PDOS: (a) Pt both top and bottom (B+C sites); (b) Pt bottom layer only (C sites).

**Table S6** The parameters related to Eq. (S18) for Pt, Sr, Ti, O photoelectron emission at 3 keV. [8, 9] $J$: the quantum number of the total angular momentum of the electron, $\sigma$: total cross-section of a subshell, $\beta$: dipole parameter, $N$: electron number, $\sigma'$: total cross-section of a shell, $\kappa$: correction factor for $\sigma'$. Other than the last three collumns, the numbers in this table are for isolated neural atoms.

| Atom | Subshell | $J$ | $2J+1$ | $\sigma$ [kb] | $\beta$ | Shell | $N$ | $\sigma'$ [kb] | | $\kappa$ | $\sigma'\kappa$ | |
|---|---|---|---|---|---|---|---|---|---|---|---|---|
| Pt | Pt $4f_{7/2}$ | 7/2 | 8 | 11.585 | 0.972 | Pt 4f | 14 | 39.98 | $\sigma_f':\sigma_d':\sigma_s'$ | 0.003 | 0.12 | $\kappa\sigma_f':\kappa\sigma_d':\kappa\sigma_s'$ |
| | Pt $5d_{5/2}$ | 5/2 | 6 | 2.893 | 1.374 | Pt 5d | 9 | 10.30 | **3.9 : 1 : 0.02** | 0.02 | 1 | **0.12 : 1 : 0.02** |
| | Pt 6s | 1/2 | 2 | 0.174 | 1.890 | Pt 6s | 1 | 0.25 | | 3.99 | 1 | |
| Sr | Sr $4p_{3/2}$ | 1/2 | 4 | 1.364 | 1.498 | Sr 4p | 6 | 5.11 | $\sigma_{4p}':\sigma_s':\sigma_{5p}'$ | 0.07 | 0.5 | $\kappa\sigma_{4p}':\kappa\sigma_s':\kappa\sigma_{5p}'$ |
| | Sr 5s | 1/2 | 2 | 0.091 | 1.998 | Sr 5s | 3 | 0.41 | **12.5 : 1 : 0** | 3.79 | 2.86 | **0.2 : 1 : 0.7** |
| | Sr $5p_{3/2}$ | 1/2 | 4 | - | - | Sr 5p | 0 | - | | - | 1.07 | |
| Ti | Ti $3p_{3/2}$ | 3/2 | 4 | 0.966 | 1.070 | Ti 3p | 6 | 3.00 | $\sigma_p':\sigma_d':\sigma_s'$ | 0.16 | 0.47 | $\kappa\sigma_p':\kappa\sigma_d':\kappa\sigma_s'$ |
| | Ti $3d_{3/2}$ | 3/2 | 4 | 0.022 | 0.453 | Ti 3d | 2 | 0.02 | **9.2 : 0.05 : 1** | 3.07 | 0.05 | **0.3 : 0.03 : 1** |
| | Ti 4s | 1/2 | 2 | 0.109 | 1.991 | Ti 4s | 2 | 0.32 | | 4.81 | 1.56 | |
| O | O 2s | 1/2 | 2 | 0.282 | 1.974 | O 2s | 2 | 0.84 | $\sigma_s':\sigma_p'$ | 1.78 | 1.49 | $\kappa\sigma_s':\kappa\sigma_p'$ |
| | O $2p_{3/2}$ | 3/2 | 4 | 0.019 | 0.282 | O 2p | 4 | 0.02 | **34.0 : 1** | 3.25 | 0.08 | **18.6 : 1** |